\documentclass{IEEEmce}

\usepackage[colorlinks,urlcolor=blue,linkcolor=blue,citecolor=blue]{hyperref}
\usepackage{cite}
\usepackage{graphicx}
\usepackage{subfigure}
\usepackage{url}
\usepackage{siunitx}
\usepackage{gensymb}
\usepackage{textcomp}
\usepackage[numbers,sort&compress]{natbib}
\usepackage{amsmath}
\usepackage{upmath}

\jvol{XX}
\jnum{XX}
\paper{8}
\jmonth{May/June}
\publisheddate{00 xxxx 0000}
\currentdate{00 xxxx 0000}
\jname{IEEE Consumer Electronics Magazine}
\pubyear{2024}
\doiinfo{MCE.2024.Doi Number}

\begin{document}

\sptitle{Department: Head}
\editor{Editor: Name, xxxx@email}

\title{Experimental Studies of Metaverse Streaming}

\author{Haopeng Wang}
\affil{University of Ottawa}

\author{Roberto Martinez-Velazquez}
\affil{University of Ottawa}

\author{Haiwei Dong}
\affil{Huawei Technologies Canada}

\author{Abdulmotaleb El Saddik}
\affil{University of Ottawa}

\markboth{Department Head}{Paper title}

\begin{abstract}
Metaverse aims to construct a large, unified, immersive, and shared digital realm by combining various technologies, namely XR (extended reality), blockchain, and digital twin, among others. This article explores the Metaverse from the perspective of multimedia communication by conducting and analyzing real-world experiments on four different Metaverse platforms: VR (virtual reality) Vircadia, VR Mozilla Hubs, VRChat, and MR (mixed reality) Virtual City. We first investigate the traffic patterns and network performance in the three VR platforms. After raising the challenges of the Metaverse streaming and investigating the potential methods to enhance Metaverse performance, we propose a remote rendering architecture and verify its advantages through a prototype involving the campus network and MR multimodal interaction by comparison with local rendering.

\end{abstract}

\maketitle

\enlargethispage{10pt}

\section{Introduction}
As Facebook changed its name to Meta, the concept of the Metaverse has gained more attention and is becoming increasingly popular in the industry and academia. The Metaverse provides a high-fidelity virtual scene and enables real-time interaction between users and the Metaverse, which is regarded as the next generation of the internet and social media. Zuckerberg believes the Metaverse will be the successor of the mobile internet and reach 1 billion users within the next ten years \cite{buana2023metaverse}.

Many technology giants, such as Meta, Microsoft, Google, HTC, Apple, and Huawei have been working on the research and development of Metaverse key technologies for a few years \cite{10.1145/3517745.3561417}. Recently, the announcement of the upcoming Apple Vision Pro headset would further promote the development of Metaverse. The Metaverse is a virtual-real fusion world where people are enabled to use their avatars to work, play, and communicate with each other. 
To provide an immersive, interconnected and unified virtual environment, the Metaverse must seamlessly incorporate various cutting-edge technologies, such as XR (extended reality), digital twin, blockchain, 5G and AI \cite{9944868,9913665}.
\begin{itemize}
\item \textbf{XR} refers to a combination of different immersive technologies that connect our physical world with digital environments, including VR (virtual reality), AR (augmented reality), and MR (mixed reality).  Users are able to experience the Metaverse in an immersive and interactive manner, enhancing their quality perceptions and engagement with the Metaverse.
\item \textbf{Digital twin} is a virtual representation of a physical entity. Through real-time synchronization between digital twins and their physical counterparts, digital twins facilitate monitoring, simulation, analysis, and enhancement of all physical entities. Digital twins enhance the richness and interactivity of Metaverse, making it more dynamic and interconnected.
\item \textbf{Blockchain} is a shared, immutable ledger that records committed transactions to facilitate the tracing and securing of digital assets in a commercial network. In the Metaverse, blockchain creates secured and decentralized systems, where digital assets can be truly owned by users through blockchain.
\item \textbf{5G} provides Metaverse with faster speeds, lower latency, larger capacities, and better connectivity across various devices and applications. 5G often works with edge computing, bringing computational resources closer to users. This combination further reduces latency, enabling faster data processing and responsiveness for the Metaverse.
\item \textbf{AI} facilitates the generation of content and enables the Metaverse to make intelligent decisions. In addition, it analyzes users' preferences and behaviors in the Metaverse to create personalized avatars and provide intelligent recommendations for goods or information to users. Moreover, it can create lifelike and intelligent virtual characters, which enhances the interactivity and overall realism of the Metaverse. 
\end{itemize}

With the rapid development of various technologies and the iteration of terminal devices, the construction and evolution of the Metaverse may far exceed people's expectations. Entertainment, fashion, education, gaming, and socializing in the Metaverse are already on the rise. In the Metaverse, users can travel around the world and buy clothing and accessories using digital currency. Metaverse also offers many amazing opportunities for enterprises. Companies in industries, such as real estate and e-commerce, can showcase product demos, reaching a wide audience and achieving incredible brand engagement \cite{10.1016/j.future.2023.02.008}. 

The goal of this article is to delve into network protocols and assess the performance of Metaverse streaming across various Metaverse platforms and devices in the wild infrastructure. In contrast to traditional communication ways that may rely on a single medium, the Metaverse utilizes various media types to create an immersive and interactive experience. By exploring Metaverse streaming through the perspective of multimedia communication, we can conduct a thorough analysis to discover distinctive features of the Metaverse streaming and provide valuable insights for the future development of the Metaverse. 

To achieve a comprehensive and precise analysis of Metaverse streaming, we conduct four experiments on four different Metaverse platforms, utilizing a range of devices including mobile phones, computers, and three popular XR devices: the VR  headset Oculus Quest 2, HTC Vive and the MR headset HoloLens 2. We first analyze the network traces collected from various scenarios in three VR platforms based on the open Internet environment to investigate the protocols and performance of Metaverse streaming. Furthermore, we discuss the challenges that the current Metaverse streaming is facing and investigate the potential solutions to enhance the performance of the Metaverse. Subsequently, we propose a remote Metaverse streaming architecture prototyped by a campus network and HoloLens 2 MR device. The experimental results show that remote rendering holds significant promise for Metaverse streaming.

\section{Network Trace Case Study I: \\ VR Vircadia}

\subsection{Vircadia}
Vircadia \footnote{\url{https://vircadia.com/}} is an open-source Metaverse platform that enables users to easily create 3D virtual environments that can be explored through a desktop application that supports commercial-grade VR headsets. It makes use of a distributed server architecture that facilitates effective scaling and load balancing. The domain server, assignment clients, and Interface client are important elements of its architecture. The domain server coordinates communication amongst clients, which takes care of entity management, physics, and audio. Contrarily, the Interface client links users to the virtual world. In order to deliver real-time, high-quality audio and video streaming, Vircadia makes use of WebRTC. Additionally, the system combines protocols like HTTP, HTTPS, and WebSocket to ensure seamless and secure communication between its various components.

\subsection{Experiment Setup}
\begin{figure}[htbp] 
\centering
\includegraphics[width=0.49\textwidth]{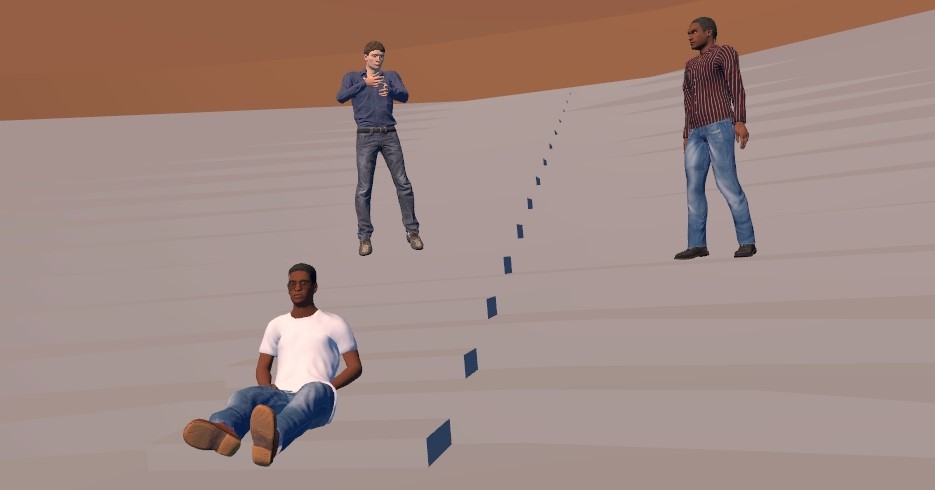}
\caption{Users in the Vircadia are having a conversation.} 
\label{vir-1}
\end{figure}

In this case study, a Vircadia instance that hosts a 3D virtual environment is deployed on a server in Ottawa. Four concurrent clients (two Oculus Quests and two computers) from Ottawa and Toronto are connected to the Vircadia server. The task assigned to the participants is to initiate a conversation on a topic of their choosing, as shown in Fig.\ref{vir-1}. Interactions between participants and with the Metaverse are moving inside the Metaverse, performing different actions, and engaging in conversations with other users. 

The VR headset used in our work is Oculus Quest 2 developed by Facebook \cite{holzwarth2021comparing}. It is a standalone VR headset that does not require a computer or a smartphone to use. The Quest 2 features a high-resolution display with a resolution of 1832 x 1920 pixels per eye, a refresh rate of 90Hz, and a field of view of 90 degrees. It is powered by a Qualcomm Snapdragon XR2 processor. Additionally, it comes with two Touch controllers for a more immersive experience.

\subsection{Network Trace Analysis}

\begin{figure}[htbp] 
\centering
\includegraphics[width=0.49\textwidth]{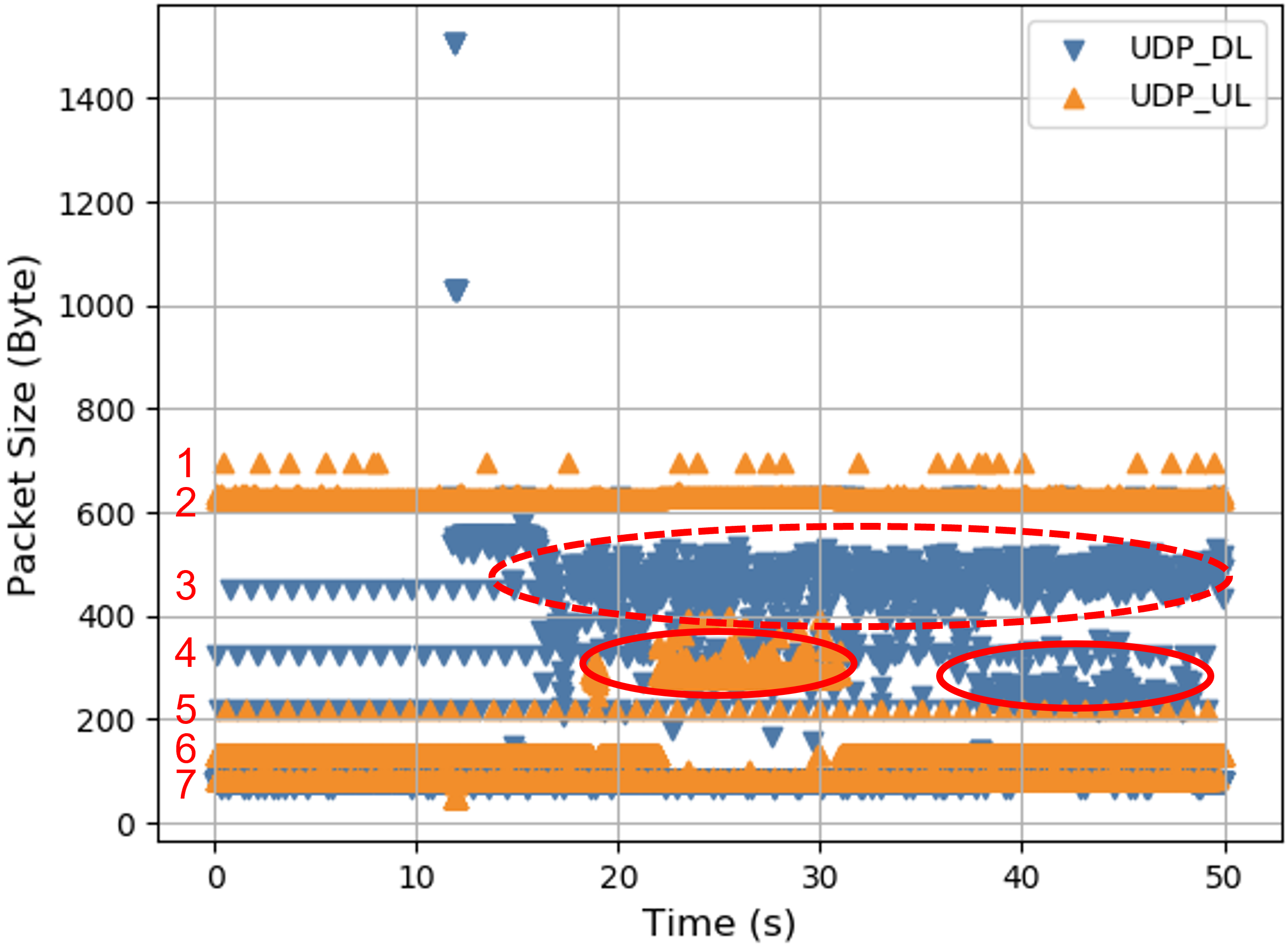}
\caption{Packets captured after uses enter the Vircadia Metaverse. The data stream is divided into the UDP uplink stream and the UDP downlink stream.} 
\label{var-1}
\end{figure}

The whole data stream can be split into an uplink (UL) stream from the client to the server and a downlink (DL) stream from the server to the client. 
In the Vircadia, the connection between clients and the server starts with the WebSocket protocol. After the Metaverse model is loaded and the connection is established, all data is streamed over UDP. The packet distribution captured in a client of a user (user 1 with computer) is shown in Fig. \ref{var-1}. User 1 stays in the Metaverse alone until user 2 (user with Oculus Quest 2) joins around 12 seconds followed by a conversion between them. It can be seen that the distribution of the packets shows strong periodicity and regularity before user 2 joins the Metaverse. Vircadia utilizes different ports for various signals, which are shown as different flows with various size levels (flow 1-7). Since the content of the Metaverse is loaded while the user accesses the Metaverse and seldom changes, the streamed packets are mainly used for connection, synchronization, acknowledgment, and interaction. 

During 12-15 seconds, user 1 receives a burst DL traffic which is the new content generated by the avatar of user 2. After that, due to the interaction with user 2, user 1 is constantly receiving DL traffic with a pattern of massive and dense packets, as shown in the red dashed circle. When the voice signal of user 1 is generated, the traffic of flow 6 increases suddenly and keeps changing according to the voice signal. The packets shown in two red circles are DL and UL voice signals respectively. All data are exchanged in a continuous and massive way with most packets sized below 700 bytes except some packets during the new content loading.

\begin{figure}[htbp] 
\centering
\includegraphics[width=0.49\textwidth]{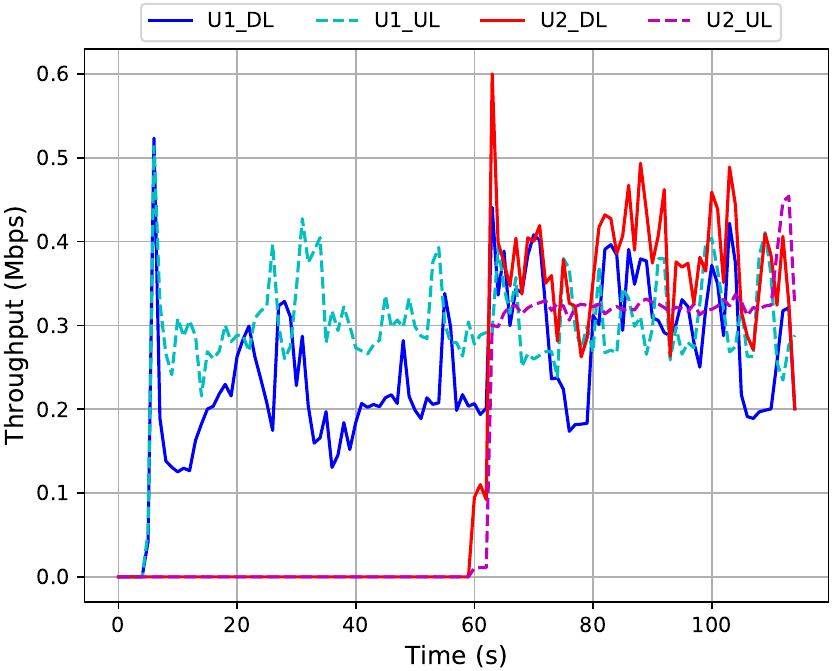}
\caption{DL and UL network throughput of Vircadia. U1 and U2 indicate the throughput of user1 and user2, respectively.} 
\label{var-2}
\end{figure}

The DL and UL throughput of the Vircadia for user 1 and user 2 are shown in Fig. \ref{var-2}. When a desktop user joins the Metaverse, a burst occurs due to the high bandwidth requirement of loading content. The UL and DL link requires similar throughput whereas the UL traffic is higher. After user 2 wearing Quest 2 joins the Metaverse, both throughputs of UL and DL keep the same level. Before user 2 joins the Metaverse, the average DL throughput of user 1 is about 0.21 Mbps, while the average UL is about 0.31 Mbps. When user 2 joins, the average DL throughput of user 1 is about 0.3 Mbps, while the average UL is about 0.31 Mbps. The average DL throughput of user 2 is about 0.37 Mbps, while the average UL is about 0.32 Mbps. The bandwidth requirement of both is similar. The network traces from different devices show a similar pattern. When a new user joins the Metaverse, the existing user will receive more DL traffic. When two users exist in the Vircadia, both have a similar throughput of 0.6 Mbps. Therefore, the total throughput is the linear addition of multiple users.

\section{Network Trace Case Study II: \\ VR Mozilla Hubs }
\subsection{Mozilla Hubs}
Mozilla Hubs \footnote{\url{https://hubs.mozilla.com/}} is an open-source Metaverse platform that enables users to explore 3D virtual worlds offering a highly immersive experience. It is web-based and supports several platforms, namely VR headsets that support web browsers, personal computers, and mobile devices. Mozilla Hubs leverages WebRTC to offer support for voice, video, and audio traffic. Furthermore, Mozilla Hubs employs a client-server architecture, with the server handling tasks such as room state management, user authentication, and content moderation. The platform is built on WebXR to provide support for VR headsets. To optimize network performance and minimize latency, Mozilla Hubs employs adaptive bitrate streaming and leverages Content Delivery Networks (CDNs) for distributing assets. Consequently, Mozilla Hubs offers a highly accessible and efficient virtual collaboration platform. Mozilla Hubs supports more interactions than Vircadia. In Mozilla Hubs, participants can share their desktop screens with other participants, and video, audio, and PDF files can be shared in real-time, allowing all participants to have synchronized views of the content. This versatility allows us to present a more interesting use case.
\subsection{Experiment Setup}
\begin{figure}[htbp] 
\centering
\includegraphics[width=0.49\textwidth]{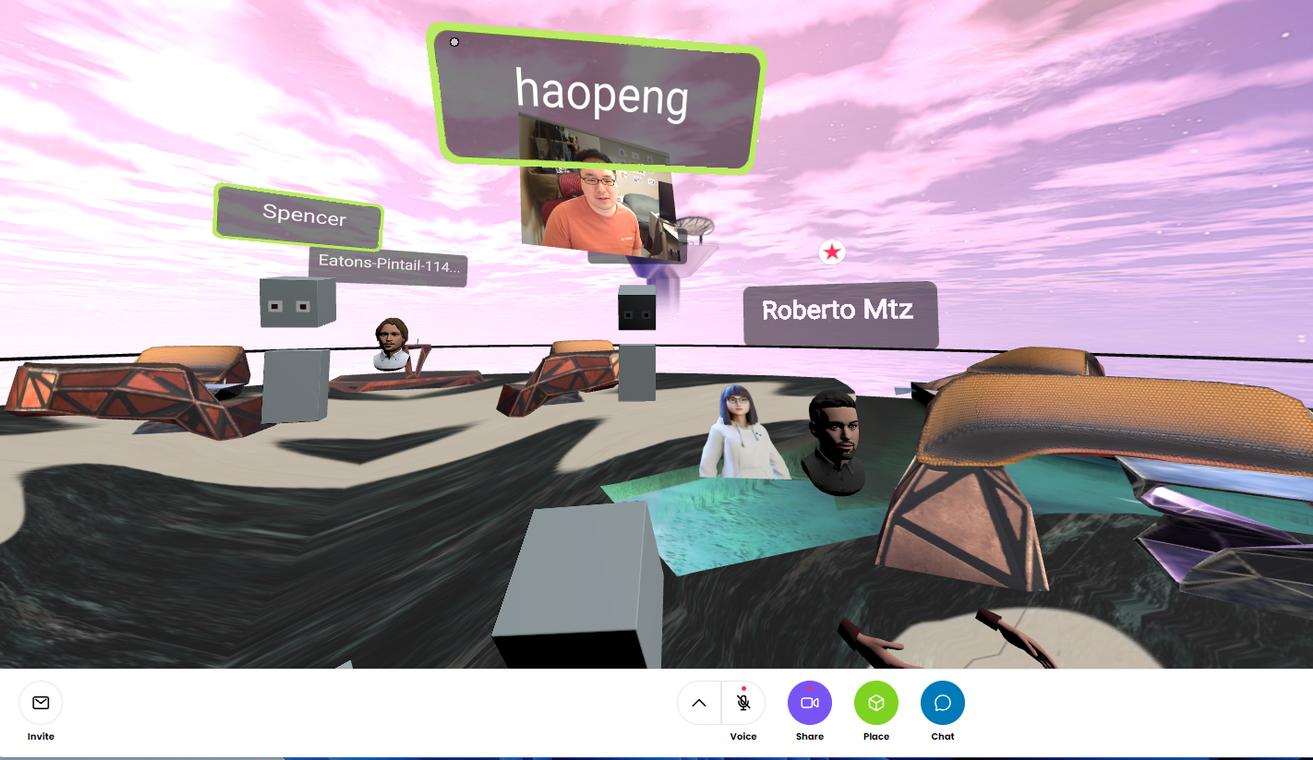}
\caption{Users in the Mozilla Hubs are having a seminar.} 
\label{hubs}
\end{figure}

In this use case, the Mozilla Hubs server is also deployed in Ottawa. Eight clients, including two Oculus Quests, three desktops, and three smartphones from different countries (Canada and China) are connected to the server, as shown in Fig.\ref{hubs}. A seminar is held in the Mozilla Hubs, where one user makes a presentation to the other users. Participants are allowed to experience the Mozilla Hubs functionalities mentioned in the above section. As a result, various data modalities can be collected, including audio, video, document, and live streaming.

\subsection{Network Trace Analysis}
\begin{figure}[htbp] 
\centering
\includegraphics[width=0.49\textwidth]{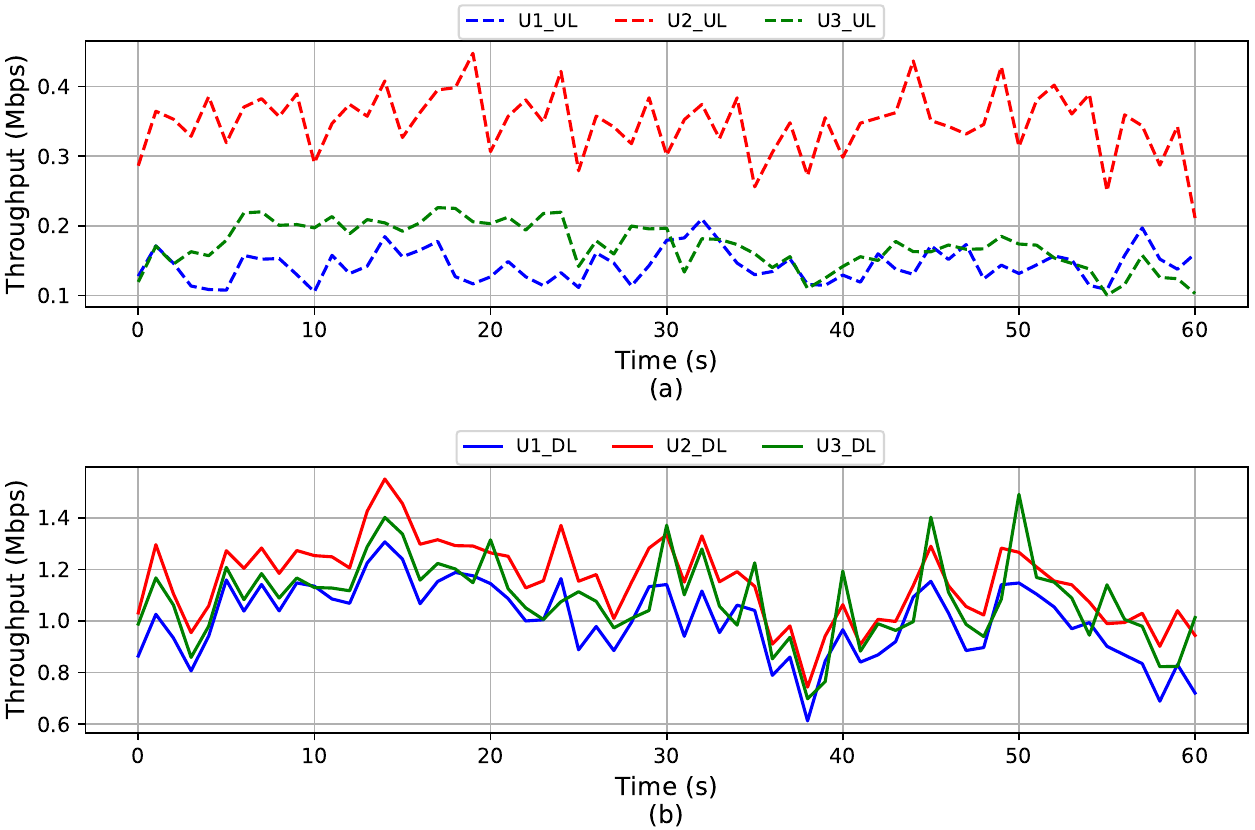}
\caption{Network throughput of three users in Mozilla Hubs. (a) UL throughput (b) DL throughput.} 
\label{hub-3}
\end{figure}

Similar to the Vircadia, network traces from different devices in Mozilla Hubs also present a similar pattern. Therefore, we present network traces of three clients in this section. The DL and UL network traces of three clients (user 1 with Oculus Quest in Ottawa, Canada, user 2 with a desktop in Montreal, Canada, and user 3 with a smartphone in Beijing, China) for a duration of 60 seconds are shown in Fig. \ref{hub-3}. All users have identical settings in Mozilla Hubs to accurately evaluate the impact of distance. Due to the same settings, different users experience similar throughputs. The Mozilla Hubs also presents a low bandwidth requirement similar to the Vircadia. The average UL throughputs for users 1, 2, and 3 are 0.14 Mbps, 0.35 Mbps, and 0.17 Mbps, respectively, while the corresponding average DL throughputs are 1.0 Mbps, 1.16 Mbps, and 1.08 Mbps. As user 2 is delivering a speech during the seminar, the UL traffic is higher than that of others. Users exhibit diverse traffic patterns in terms of UL throughput due to their various actions, while maintaining similar traffic patterns in DL throughput.

As network latency plays a crucial role in shaping the user experience, we additionally measure the latency experienced by the three users across different cities. The latency for the three users is about 17.5 ms, 18.4 ms, and 146.8 ms, corresponding to the cities of Ottawa, Montreal, and Beijing, respectively. It is evident that latency significantly increases when a user is located in a distant place, as the data transmission takes more time in such instances. The distance between the client and the Metaverse server has a large impact on latency, but has a negligible effect on network traffic, which mainly depends on the user's configuration and activities within the Metaverse.

\begin{figure}[htbp] 
\centering
\includegraphics[width=0.49\textwidth]{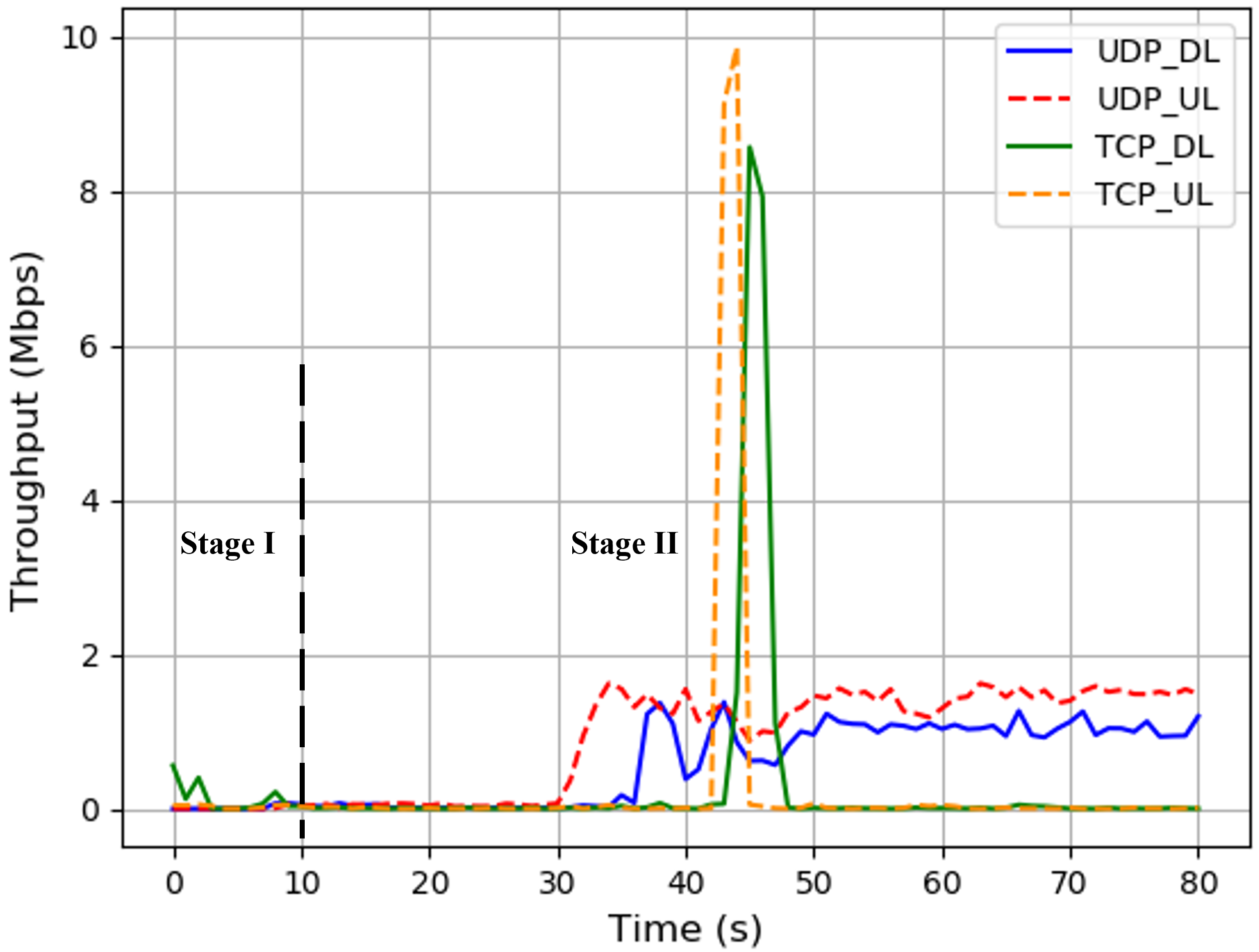}
\caption{Network throughput of Mozilla Hubs. DL and UL for TCP and UDP are shown respectively.} 
\label{hub-1}
\end{figure}

Since Mozilla Hubs is a website platform, HTTPS with TLS is used at the application layer for secure communication. The detailed traffic captured from a user's client (user 1) is shown in Fig. \ref{hub-1}. A menu page for configuring the Metaverse is loaded when user 1 accesses the website before entering the Metaverse. Simultaneously, the Metaverse model data is cached in the browser, which causes a TCP\_DL burst between 0 and 4 seconds, as shown in stage I of Fig. \ref{hub-1}. The second TCP\_DL burst occurs between 7 and 10 seconds due to the joining of user 2. TCP is used exclusively before users enter the Metaverse. 

After entering the Metaverse, both TCP and UDP are used, as shown in the second stage of Fig. \ref{hub-1}. UDP is used for streaming real-time data, such as voice, and live video streaming, which requires low latency. From 10 to 30 seconds, only voice signals are generated due to the conversation between two users and transmitted over UDP. The voice signals have a small data volume, approximately 45 Kbps. Subsequently, UDP\_DL and UDP\_UL suddenly surge to high levels (about 1.3 Mbps) as both users turn on their webcams. TCP is primarily used for data that is more important and less sensitive to latency, such as updated Metaverse content, control signal, and acknowledgments. The Mozilla Hubs supports uploading and sharing various multimedia files, such as documents, videos, and audio, all of which are streamed using TCP. As shown in stage II of Fig. \ref{hub-1}, two TCP bursts, TCP\_UL and TCP\_DL, are generated due to file upload. The file is first uploaded to the server, and then downloaded and displayed to users in the Metaverse, resulting in a peak of about 10 Mbps. The peak of the burst depends on the size and number of uploaded files. During the 10 minutes of streaming, 24 bursts of over 50 Mbps occur. The highest observed burst peak on the server reaches up to 570 Mbps for 8 users in the Mozilla hubs. The average throughput is 8 Mbps; however, the average throughput without the burst is about 1.7 Mbps. Additionally, it can be seen that bursts occur frequently due to uncertainty about user activity.

Additionally, we observed that Mozilla Hubs has fewer regular periodic packets compared to other platforms. We speculate that both TCP and UDP are used for Metaverse streaming. TCP utilizes corresponding acknowledgment packets, whereas WebRTC over UDP relies on RTP and RTCP. RTCP is responsible for monitoring network conditions and providing feedback information to the server.

\section{Network Trace Case Study III: \\ VRChat }
\subsection{VRChat}
VRChat \footnote{\url{https://hello.vrchat.com/}} is a popular and free Metaverse platform launched by VRChat Inc. in 2017, which supports multiple users and VR headsets, such as HTC Vive, Oculus Quest, computers, and so on. Users have the ability to generate and share their personalized avatars, 3D models, and virtual worlds, fostering a diverse and imaginative community enriched with a broad spectrum of virtual settings and characters.
\subsection{Experiment Setup}

\begin{figure}[htbp] 
\centering
\includegraphics[width=0.48\textwidth]{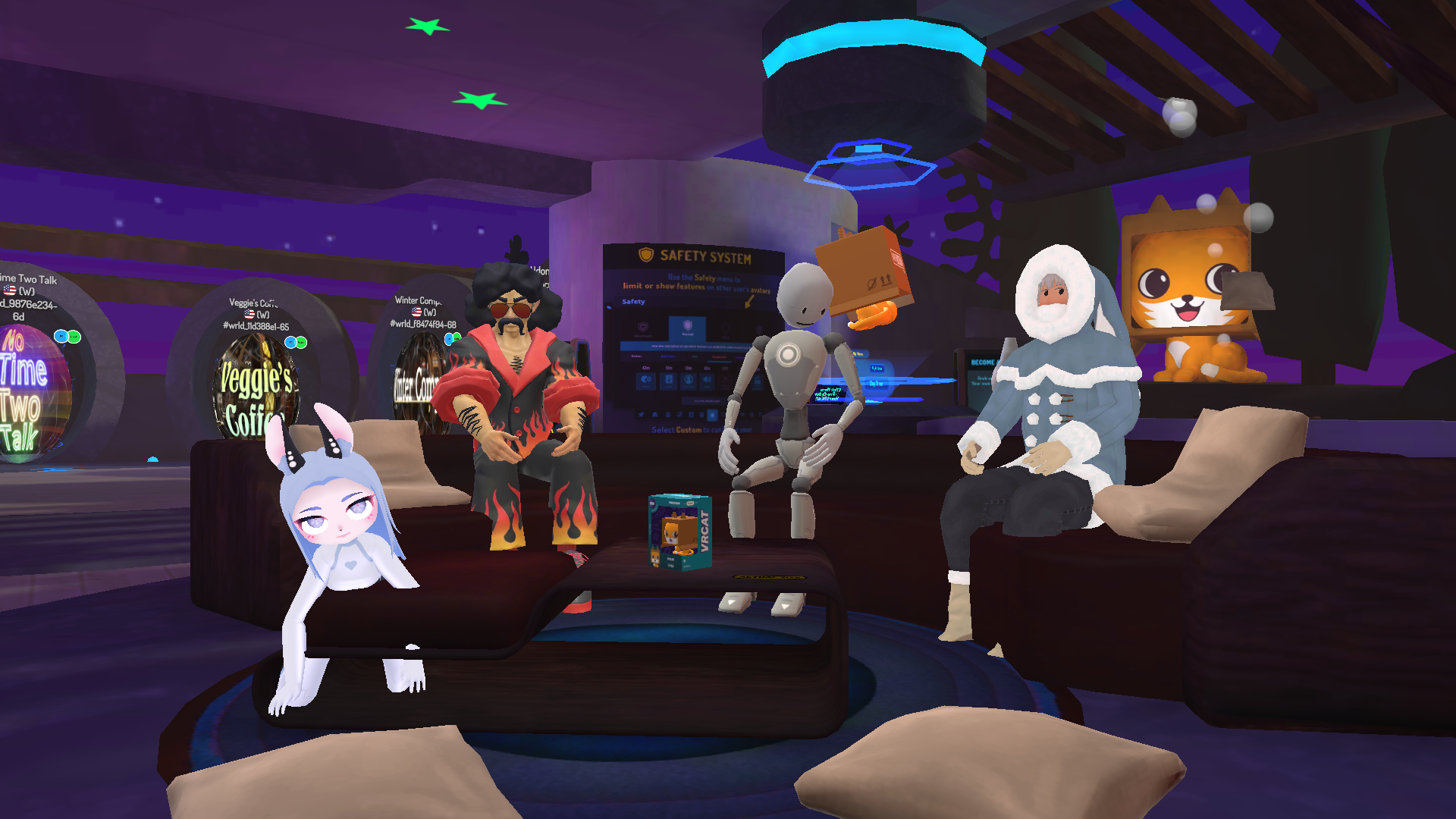}
\caption{Users in VRChat are having a conversation.} 
\label{vrchat}
\end{figure}

In this use case, four clients, including one Oculus Quest, one HTC Vive and two desktops from Ottawa and Montreal, are connected to the VRChat server. HTC Vive is a VR headset with a resolution of 1080 x 1200 pixels per eye, a refresh rate of 90 Hz and a 110-degree field of view. A meeting is held in the VRChat Metaverse, as shown in Fig.\ref{vrchat}, where users can perform various activities and discuss with others.

\subsection{Network Trace Analysis}
\begin{figure}[htbp] 
\centering
\includegraphics[width=0.5\textwidth]{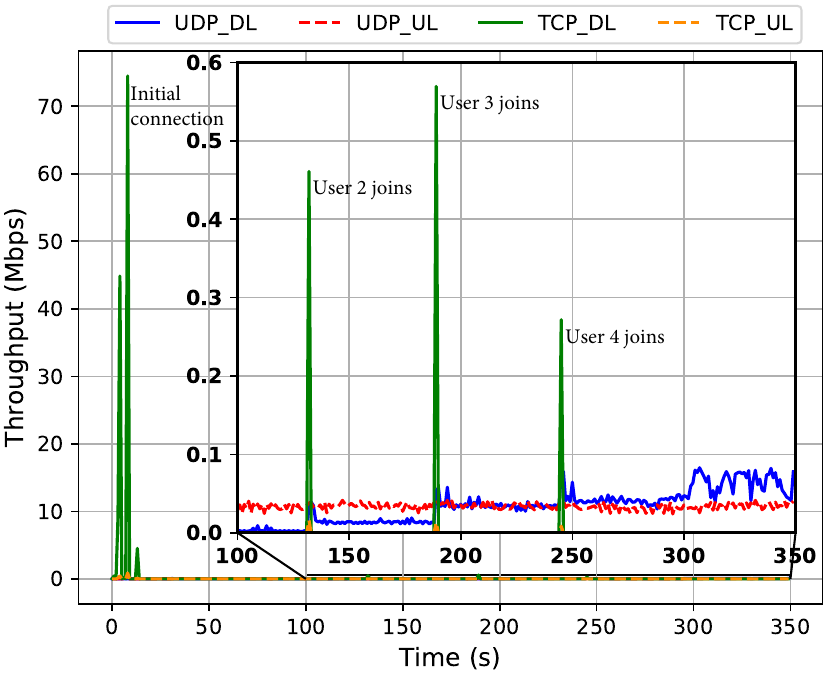}
\caption{Network throughput of user 1 in VRChat. DL and UL for TCP and UDP are shown respectively.}
\label{vrchat-1}
\end{figure}

Since all users exhibit a similar network trace pattern, we showcase a 350-second network trace of a user (user 1) in VRChat, as shown in Fig. \ref{vrchat-1}. The HTTPS with TLS and UDP are used in the VRChat. Similar to Vircadia and Mozilla Hubs, VRChat also consists of two stages: connection and transmission. In these two stages, two servers are utilized: one for connection located in Montreal and supported by Amazon, and the other for data transmission situated in San Francisco and supported by Cloudflare. During the initial connection stage from 0 to 15 seconds, the Metaverse model and additional information, including web page data and synchronization signals, are transmitted via TCP. Since the VRChat has a sophisticated scenario, the average throughput of TCP\_DL in the connection stage is approximately 8.9 Mbps with a peak of 74.5 Mbps while the average throughput of TCP\_UL is around 0.1 Mbps.

In the second stage, UDP is employed for data transmission. Since only user 1 is in the VRChat from 15 to 135 seconds, the average throughput of UDP\_DL is approximately 0.002 Mbps while the average throughput of UDP\_UL is about 0.03 Mbps. A burst is triggered upon the entry of a new user into the Metaverse, coinciding with the loading of a new avatar. As users 2, 3, and 4 join VRChat at around 131s, 188s and 244s respectively, the UDP\_DL throughput exhibits an almost linear increase. With a total of four users in the Metaverse, the average UDP\_DL throughput is around 0.04 Mbps. As user 1 engages in basic actions like walking and jumping, the UDP\_UL throughput remains stable. After 300 seconds, when one user initiates speech, the average UDP\_DL throughput increases to 0.07 Mbps. Similar to the Mozilla Hubs, users in different cities also have comparable network traces. The distance also exerts a significant impact on latency, with approximately 104 ms latency in Ottawa and about 111 ms in Montreal.

\section{Challenges in Current Metaverse Streaming}
Current Metaverse streaming typically consists of two stages: connection and transmission. The TCP is utilized for establishing a stable connection between the client and the server, whereas the UDP is employed for streaming Metaverse data. Since a Metaverse is envisioned as a vast network of interconnected virtual worlds where millions of people can interact simultaneously, the UDP is considered more suitable for Metaverse streaming due to its low latency for real-time data streaming and scalability to handle a large number of users. 

However, based on the above analysis, we find there are still several challenges in Metaverse streaming. First, the current Metaverse lacks the ability to offer users a seamless and intuitive method of interacting with the virtual world, objects, and other users \cite{10130406}. Avatars in the current Metaverse are constrained to a narrow selection of actions achievable through handheld controllers, lacking the capacity for nuanced and intricate body movements. Second, the Metaverse is currently in its infancy with low quality, therefore requiring low computational and bandwidth resources. However, the future Metaverse will demand high-quality content to provide a truly immersive experience for millions of concurrent users, necessitating higher bandwidth and computational power than what is currently observed \cite{9984845}. Last but not least, it can be seen that the three VR platforms render the Metaverse directly on the headset, resulting in significantly low throughput. This local rendering mechanism is also adopted by other popular Metaverse platforms \cite{10.1145/3517745.3561417}. However, the computational power of headsets is limited, which restricts the user experience in Metaverse.

\section{Potential Solutions}
The Metaverse aims to provide a fully immersive experience with seamless interaction, which requires low latency and high data rate. To address the aforementioned challenges, there are various promising research directions that can be explored. 
\begin{itemize}
    \item{\textbf{Bandwidth prediction:}} A primary challenge in Metaverse streaming is the high volatility of actual network, especially wireless link, which leads to high delay and low quality of experience. The future bandwidth can be estimated in advance by employing encoding networks, such as Long short-term memory (LSTM) and transformer variants, based on the bandwidth from past and present \cite{9024132}. The Metaverse could utilize the predicted results to adjust its transmission strategies in real time.

    \item{\textbf{Adaptive streaming:}} As network conditions fluctuate, the Metaverse could autonomously ascertain diverse streaming configurations or parameters, such as content quality, frame rate, and CPU/GPU resources, to adapt to these variations. Deep reinforcement learning (DRL) could enable the system to adaptively learn and implement the optimal streaming strategy \cite{10144631}. 
    
    \item{\textbf{Viewport streaming:}} Due to the characteristics of the human visual system, users only view a portion of Metaverse content within the Field of View (FoV). The content outside the viewport is either not immediately utilized or never viewed. Consequently, the user’s viewport can be transmitted at higher quality, while the area outside of the viewport is streamed at lower quality or even not streamed at all \cite{9086630}.

    \item{\textbf{Haptics:}} The current Metaverse provides only fundamental visual and auditory experiences. Haptics technology has the capability to offer users tactile feedback, encompassing sensations such as touch, pressure, and kinesthetic feedback involving motion and resistance \cite{9984845}. The integration of haptics in the Metaverse offers a more comprehensive and engaging experience.

    \item{\textbf{Remote rendering:}} From the analysis of the previous three case studies, it's evident that current Metaverse platforms rely solely on local rendering, leading to a limited quality of experience due to constrained computational resources. Hence, we propose a remote rendering solution for the Metaverse, where the rendering workload is offloaded to a remote server with robust computational resources, instead of being processed on the local clients.

\end{itemize}

\section{Outlook: Remote Rendering for Metaverse Communication}
Since Metaverse streaming heavily depends on its rendering mechanisms, we investigate the performance of local rendering on a client and remote rendering on a remote server. The former is to perform full rendering on the client, which renders Metaverse content interactively through the GPU/CPU on the client.
The latter refers to rendering Metaverse content remotely on a computing device called rendering server and streaming the rendered results to another network-connected display device called client \cite{10.1145/2719921}.

Compared with local rendering, remote rendering has the following advantages. First, it can provide a powerful rendering ability for compact Metaverse clients (e.g., headsets) with limited computational resources. Second, multiple clients can efficiently share computational resources on a rendering server. Third, remote rendering provides a cross-platform solution. Metaverse developers only need to focus on the development of Metaverse content on a remote rendering server while disregarding the differences between platforms. Last but not least, remote rendering systems can protect intellectual property since users only receive the rendering results streamed from the server.

\begin{figure}[htbp] 
\centering
\includegraphics[width=0.49\textwidth]{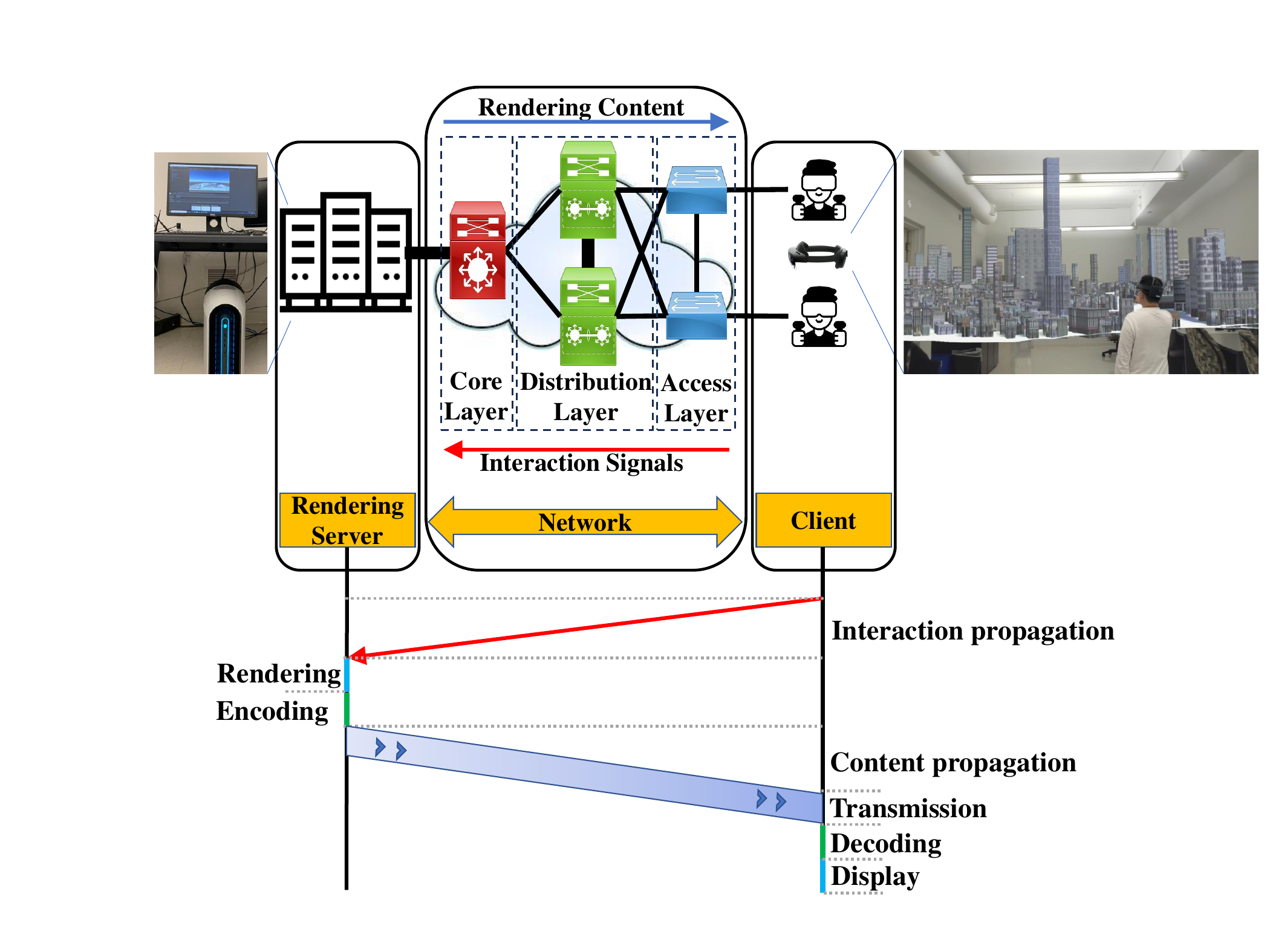}
\caption{The proposed Metaverse remote rendering system.} 
\label{framework}
\end{figure}
\subsection{Experiment Setup}
To explore the potential of remote rendering and streaming features of the Metaverse, we propose a Metaverse remote rendering system with a campus network architecture, as shown in Fig. \ref{framework}. The remote rendering system contains three components: rendering server, network, and client. The remote rendering server is used to provide high computational power to render high-quality Metaverse content while the client displays the content and enables users to interact with the Metaverse. The network connecting the client with the server in the system used in our system is a campus network, as the campus network is cost-effective and easy to implement within a geographical area, which provides a high data transfer rate for the Metaverse. The campus network is commonly deployed in enterprise network scenarios, such as VR classrooms in universities, AR surgery in hospitals, etc. The campus network has a three-layer hierarchical topology: core layer, distribution layer, and access layer. The core layer is the backbone of the campus network and connects different distribution layer devices. The main purpose of the core layer is to exchange data as fast as possible. The distribution layer is a service and control boundary between the access layer and the core layer, which defines policy and ensures stability for the campus network. The access layer is responsible for enabling clients to access the campus network. 

To evaluate the proposed remote rendering system, a desktop with a Nvidia RTX3090 GPU is used as the remote rendering server. The campus network is simulated on a desktop with the ns3 network simulator \cite{henderson2008network} to connect the server and client. In the simulated campus network, as shown in Fig. \ref{framework}, the core layer, distribution layer, and access layer are composed of a router, two routers, and two switches, respectively. A HoloLens 2 serves as the client, which is a MR headset designed by Microsoft. It has a FoV of 52 degrees diagonally and a resolution of 2048x1080 pixels (per eye) powered by 2K 3:2 light engines. Besides, it has a depth camera based on active infrared illumination to determine depth, an RGB camera to capture images or videos, two infrared cameras for eye tracking and iris recognition, and four visible light tracking cameras for head tracking and real-time visual-inertial SLAM (simultaneous localization and mapping).

The Metaverse scenario used in the proposed system is a Virtual City. The virtual city is rendered using Unity in the remote rendering server and streamed to the HoloLens 2 via the campus network. A user wearing HoloLens 2 is capable of not only viewing but also interacting with the virtual city transmitted from the rendering server in real time. We can interact with the Metaverse system with various multimodal approaches, such as hand gestures, physical movement, and voice recognition. The hand gestures of a user are captured by a camera built in the HoloLens 2, which moves or resizes the Metaverse objects, selects buildings we want to view, and even zooms into a detailed street view. The HoloLens 2 interprets physical movement through its rotation and position sensors. Using this data, it calculates the user's location and subsequently updates the Metaverse content to align with the user's movement. Voice recognition allows the user to easily interact with the Metaverse, such as resizing an object, logging into or out of Metaverse, enabling performance measurement for the Metaverse system, etc. As shown in Fig. \ref{framework}, when a user interacts with a Metaverse, interaction signals are generated and propagated to the rendering server. After receiving the interaction signals, the remote server renders new frames and encodes the frames to reduce transmission consumption. After all frames are transmitted into the network, they are propagated to the client over the network. Once the client receives all the data of the frames, it decodes the data into frames for display.

\begin{figure}[htbp] 
\centering
\includegraphics[width=0.49\textwidth]{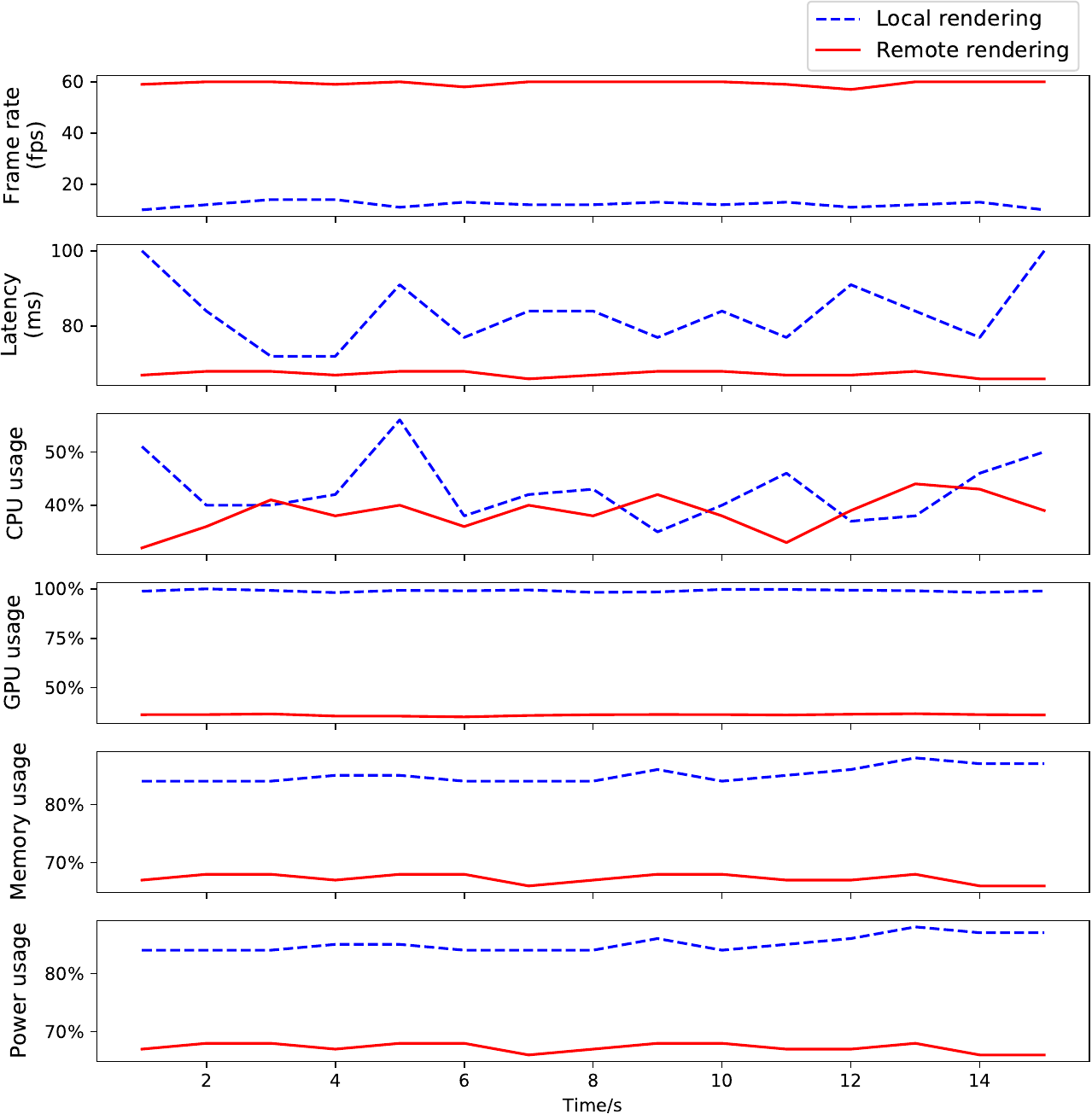}
\caption{Performance comparison between local rendering and remote rendering.} 
\label{lr}
\end{figure}

\subsection{Performance Analysis}
The evaluations of local rendering on the HoloLens 2 and remote rendering on the server of the proposed remote rendering system are conducted with the Virtual City platform and H.264 codec. As shown in Fig. \ref{lr}, the frame rate using remote rendering is 60 fps, while the frame rate with local rendering is approximately 12 fps. Although the GPU is used to render the Metaverse content (99\%) in local rendering, it still takes a lot of time to render each frame due to the computational power limit, which leads to high latency (85 ms) compared to the latency (67 ms) of remote rendering. Since GPU is mainly employed to render the content, both rendering methods have similar CPU usage (46\% for local rendering and 40\% for remote rendering). The memory usage and power usage are approximately 68\% and 78\% for remote rendering compared to 84\% and 90\% for local rendering, respectively. Since the server works on the heavy rendering workload in the remote rendering system, the computational requirement for the client is much less compared to local rendering, leading to a more compact device with a better user experience.

\subsection{Network Trace Analysis}
\begin{figure}[htbp] 
\centering
\includegraphics[width=0.49\textwidth]{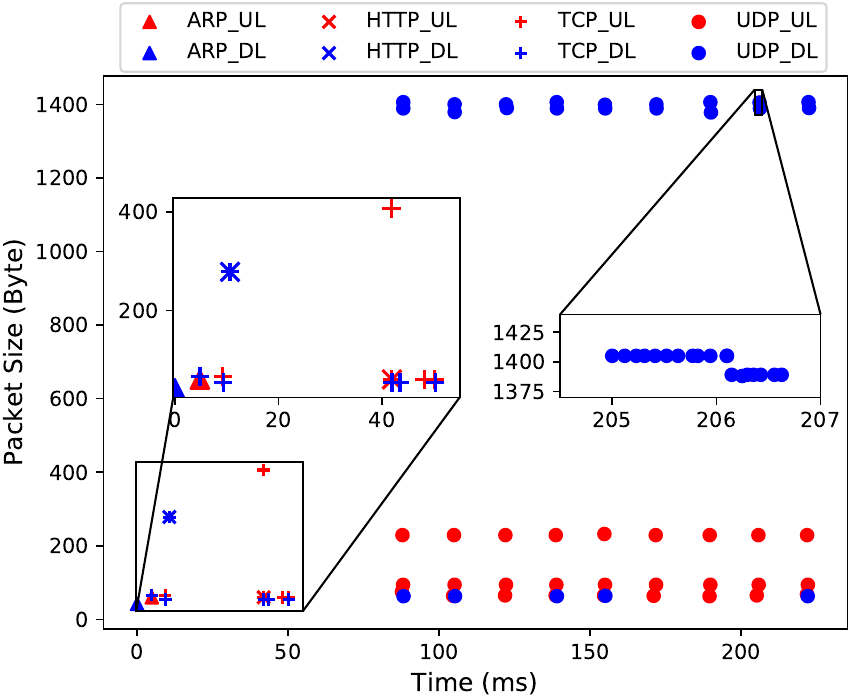}
\caption{Packets captured during connection and transmission stages in the remote rendering system.}
\label{packet}
\end{figure}
The network trace of the user is shown in Fig. \ref{packet}. Similar to the VR platforms, the remote rendering MR system also has two phases, connection and transmission, to stream Metaverse content from the rendering server to the HoloLens 2. In the connection phase, as shown in the left enlarged view of Fig. \ref{packet}, the rendering server and HoloLens 2 complete the initial connection for future data transmission by building an HTTP session. Once the connection is established, all types of Metaverse data are streamed over UDP. As Metaverse is sensitive to latency, UDP can deliver data at a faster speed compared to TCP. The UDP\_UL stream at the transmission phase is composed of synchronization information, feedback of transmitted frames, and sensory data (such as head pose tracking, hand gesture tracking, and voice recognition). Although the UDP\_DL stream also contains synchronization information, it is mainly composed of frame packet bursts which contain several UDP packets of over 1000 Bytes, as shown in the right enlarged view of Fig. \ref{packet}. There are two bursts corresponding to two frames transmitted to the left eye and right eye. Since HoloLens 2 has a frame rate of 60 fps, the average interval between successive bursts is 16.67 ms.

\section{CONCLUSION}
The Metaverse is a powerful technology providing new immersive experiences as the next-generation Internet, although it is in the preliminary stage with many scientific/technical challenges. By conducting experiments with VR platforms, we explore the network protocols used in these platforms and evaluate network performance under the current streaming method and network. The challenges for rendering high-quality Metaverse with limited computational power as well as multi-modality interaction are discussed and analyzed. Since the future Metaverse will yield high requirements for computational and bandwidth resources, some potential solutions are investigated to enhance the performance of the Metaverse. As the high computational load of Metaverse could be offloaded on a powerful remote rendering server, the article quantitatively verifies the performance improvement of the remote rendering in comparison to that of the local rendering in Metaverse streaming way.

\bibliographystyle{IEEEtran}
\bibliography{ref.bib}

\begin{IEEEbiography}{Haopeng Wang} is a Ph.D. candidate with the School of Electrical Engineering and Computer Science, University of Ottawa, Ottawa, ON, Canada. His research interests include multimedia, extended reality, and artificial intelligence. Wang received his M.Sc. degree in electrical and computer engineering from the Beijing Institute of Technology, Beijing, China. Contact him at hwang266@uottawa.ca.
\end{IEEEbiography}

\begin{IEEEbiography}{Roberto Martinez-Velazquez} is a Ph.D. candidate with the School of Electrical Engineering and Computer Science, University of Ottawa, Ottawa, ON, Canada. His research interests include Digital Twins, eHealth, multimedia, and Metaverse. Martinez-Velazquez received his M. Sc. degree in Computer Science from the CICESE in Mexico. Contact him at rmart121@uottawa.ca.
\end{IEEEbiography}

\begin{IEEEbiography}{Haiwei Dong} is a Principal Researcher with Huawei Technologies Canada, Ottawa, ON K2K 3J1, Canada. His research interests include artificial intelligence, multimedia, Metaverse, and robotics. Dong received his Ph.D. degree from Kobe University, Kobe, Japan. He is a senior member of the IEEE. Contact him at haiwei.dong@ieee.org.
\end{IEEEbiography}

\begin{IEEEbiography}{Abdulmotaleb El Saddik} is a Distinguished University Professor with the University of Ottawa, Ottawa, ON, Canada. His research interests include multimodal interactions with sensory information in smart cities. He is a Fellow of Royal Society of Canada, a Fellow of IEEE, an ACM Distinguished Scientist and a Fellow of the Engineering Institute of Canada and the Canadian Academy of Engineers. He was the recipient of the IEEE I\&M Technical Achievement Award, the IEEE Canada C.C. Gotlieb (Computer) Medal, and the A.G.L. McNaughton Gold Medal. Contact him at elsaddik@uottawa.ca.
\end{IEEEbiography}

\end{document}